\documentclass[a4paper, reprint, twoside]{revtex4-2}

\usepackage[english]{babel}
\usepackage[utf8]{inputenc}
\usepackage[labelfont=up]{subcaption}
\usepackage[x11names]{xcolor}
\usepackage{float}
\usepackage{mathtools}
\usepackage[top=1.in, bottom=1.in, left=0.5in, right=0.5in]{geometry}
\usepackage{amsthm, amssymb}
\usepackage{graphicx}
\usepackage{lipsum}
\usepackage{bm}
\usepackage[pdftex, pdftitle={Article}, pdfauthor={Author}]{hyperref}
\usepackage{wasysym}
\usepackage{enumitem}
\usepackage{comment}
\usepackage{adjustbox}
\hypersetup{colorlinks=true, citecolor=black, linkcolor=black, urlcolor=blue}

\newcommand{\upd}[1]{{\color{black} #1}}

\begin{document}

\title{Species Vulnerability and Ecosystem Fragility: A Dual Perspective in Food Webs}
\author{Emanuele Cal\`o$^1$, Giordano De Marzo$^{2, 3, 4}$ and Vito D.~P.~Servedio$^4$}
\affiliation{
$^1$Scuola IMT Alti Studi Lucca, Piazza S. Francesco, 19 - 55100 Lucca, Italy.\\ 
$^2$University of Konstanz, Universitaetstrasse 10, 78457 Konstanz, Germany.\\
$^3$Centro Ricerche Enrico Fermi, Piazza del Viminale, 1, I-00184 Rome, Italy.\\
$^4$Complexity Science Hub, Metternichgasse 8, 1030, Vienna, Austria.}
\date{\today} 

\begin{abstract}
    \noindent Ecosystems face intensifying threats from climate change, overexploitation, and other human pressures, emphasizing the urgent need to identify keystone species and vulnerable ones. While established network-based measures often rely on a single metric to quantify a species’ relevance, they overlook how organisms can be both carbon providers and consumers, thus playing a dual role in food webs. Here, we introduce a novel approach that assigns each species two complementary scores—an importance index quantifying their centrality as carbon source and a predatory index capturing their vulnerability. We show that species with high importance index are more likely to trigger co-extinctions upon removal, while high-robustness index species typically endure until later stages of collapse, in line with their broader prey ranges. On the other hand, low \upd{robustness} index species are the most vulnerable and susceptible to extinctions. Tested on multiple food webs, our method outperforms traditional degree-based analyses and competes effectively with eigenvector-based approaches, while also providing additional insights. 
    This scalable and data-driven approach, relying solely on interaction data, provides a cost-effective tool that complements expert classifications for prioritizing conservation efforts.
\end{abstract}
\maketitle
\section{Introduction}
    \noindent
    The preservation of ecosystems is one of the most pressing challenges in contemporary ecology. A key aspect of this challenge lies in understanding the consequences of species loss and identifying the most vulnerable species. These tasks are increasingly urgent due to escalating anthropogenic pressures, such as climate change and over-exploitation, which severely disrupt ecosystems \cite{petchey1999environmental, harmon2009species, blanchard2015rewired, morris2016deforestation, martinez2016anthropogenic, bartley2019food, nagelkerken2020trophic, gomes2024marine}. Species extinctions often lead to cascading effects throughout the ecosystem, triggering secondary extinctions that amplify the initial damage \cite{dunhill2024extinction, keyes2024synthesising}. To mitigate these impacts, it is essential to quantify species' "relevance" and "robustness," guiding conservation policies to prioritize the preservation of fundamental and fragile species.
    
    Ecosystems, like many complex systems, are characterized by non-linear interactions and intricate structures \cite{fussmann2002food, pillai2011metacommunity}, where each component contributes to overall functionality. Similar dynamics are observed in other domains, such as interbank loan markets or supply chains, where the removal of a single bank or firm can significantly disrupt economic output \cite{karimi2016cascades,diem2022quantifying, stangl2024firm}. In this context, network science has emerged as a powerful tool for studying such systems. By representing species as nodes and their interactions—such as mutualistic, host-parasite, or trophic relationships—as links, networks provide a mathematical framework for analyzing ecological systems  \cite{eklof2013dimensionality, dominguez2015ranking}. Among these networks, food webs, which depict "who consumes whom" interactions, have been extensively studied to understand ecosystem dynamics \cite{dunne2002food, garlaschelli2003universal, krause2003compartments, otto2007allometric, allesina2008general, ghoshal2011ranking, allesina2015predicting, keyes2021ecological}.
    Network-based measures, particularly those derived from PageRank or eigenvector centrality \cite{page1999pagerank, bryan200625, thurner2018introduction}, have proven effective in quantifying species' relevance within food webs. These metrics identify species whose removal would have the most significant impact on biodiversity \cite{allesina2009googling, mcdonald2016using}. This approach is especially useful for large food webs, where exhaustive simulations of all extinction scenarios are computationally infeasible. However, while these methods excel at identifying critical species, they often overlook those that are highly vulnerable. For instance, some species may play a minor role in the food web economy but remain extremely susceptible to extinction due to specialized diets or habitat requirements—pandas being a notable example.

    \upd{While network science has provided numerous metrics for quantifying species' roles in food webs, including measures of centrality and structural roles \cite{estrada2007characterization, luczkovich2003defining}, and the concept of keystone species has been explored through various network approaches \cite{jordan1999reliability}, our methodology offers a distinct perspective by explicitly and simultaneously quantifying two complementary dimensions for each species: the importance index (measuring its centrality as a carbon source and its relevance within the food web) and the robustness index (capturing its predatory ability and resilience to extinction). Unlike single-index measures that characterize different aspects of species' network roles \cite{cirtwill2018review, rocchi2017key}, this bi-dimensional framework provides a direct and separate assessment of these two fundamental ecological facets for each species, building upon the idea that considering both a species' dependence on resources and its impact on its prey is crucial for a comprehensive understanding of its ecological role, as also explored through combinations of centrality indices \cite{gouveia2021combining}. This bi-dimensional characterization then allows us to visualize the food web's structure, where each species is represented as a point on this plane.}
    Our findings reveal that highly important species often drive significant extinction cascades when removed, whereas low-\upd{robustness} index species exhibit limited resilience to such events. This methodology provides a more nuanced understanding of both critical and vulnerable species within ecosystems. 
    By capturing the complex interplay between predation and survival in food webs, our approach \upd{offers a valuable, ecologically-informed perspective that could contribute to the broader framework of conservation strategies, potentially enhancing the prioritization of biodiversity protection efforts.}
    
    \begin{figure*}[t]
            \centering
            \includegraphics[width=0.9\textwidth]{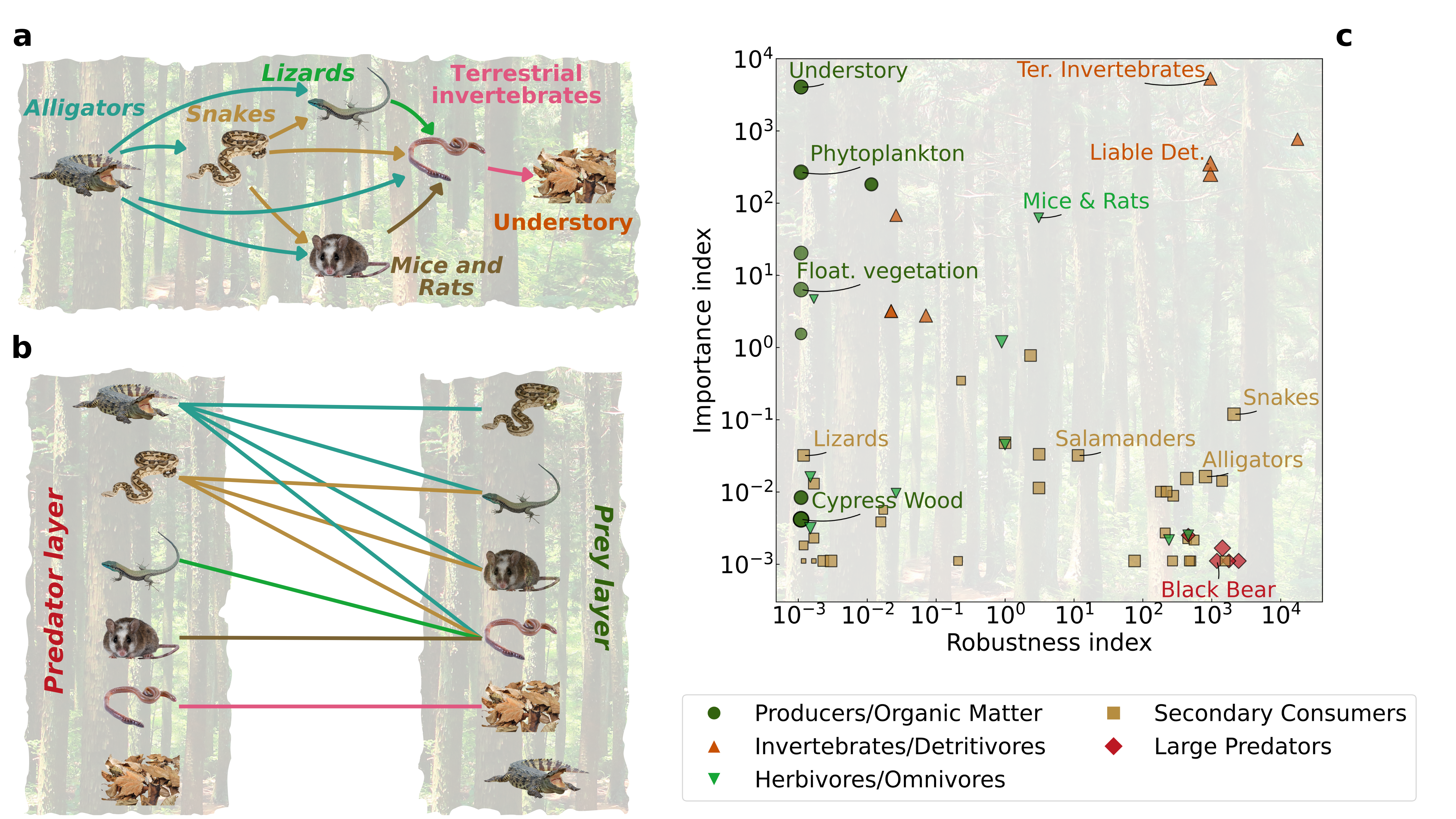}
            \caption{\textbf{Cypress food web: network representations and \upd{robustness} index-importance index plane.} 
            \textbf{a,} Directed monopartite subnetwork illustrating predator-prey relationships.
            \textbf{b,} Equivalent undirected bipartite representation of the subnetwork.
            \textbf{c,} \upd{Robustness} index-importance index plane depicting species' ecological roles. Points represent individual species, with colors indicating functional categories and sizes proportional to species biomass. Selected species are labeled for reference. This visualization reveals the dual roles of species as both predators and prey within the ecosystem.}
            \label{fig:cypress}
    \end{figure*}
        
\section{Results}
    \subsection{The \upd{Robustness} index and Importance index of Species}
\noindent
        Quantifying species' roles and impacts within ecosystems presents a significant methodological challenge in ecological research. Traditional network-based methodologies approach this task by looking at the food web formed by the species. This consists of a directed network, where nodes are the species, while edges signify carbon \upd{flow} between these species. 
        \upd{As predation represents a major pathway for carbon flow between species, these networks are frequently termed prey-predator networks.}
        We show in Fig.~\ref{fig:cypress}a a schematized food web, where we adopt the convention of arrows going from predators to prey. Once a food web has been reconstructed, the standard approach consists in applying a centrality measure to rank the species. For instance, eigenvector centrality has proven very effective in determining which species have a stronger potential to cause chain extinctions if removed from an ecosystem \cite{allesina2009googling}. These measures, however, are unable to assess which species are most in danger during these extinction cascades \upd{and may not fully capture species' vulnerability or the nuances of direct and indirect interactions (see also mixed trophic impact analysis \cite{ulanowicz1990mixed} and loop analysis \cite{bodini1994qualitative})}.
        
        To address this limitation, we introduce a novel algorithmic approach that quantifies species' dual roles through two distinct measures: the importance index and the \upd{robustness} index. The importance index measures how important in the food web a species is, while the \upd{robustness} index assesses a species' predatory capabilities and adaptability within the food web. The importance index of a species is high if it serves as prey for multiple species with a low \upd{robustness} index. On the other hand, the \upd{robustness} index evaluates a species' predatory efficiency by assessing both the quantity and the importance index of its prey species. 
        Species that prey upon diverse taxa, particularly those of low importance index, demonstrate high predatory capacity and consequently receive high \upd{robustness} index values. 
        Mathematically, we formulate this framework as a bi-dimensional non-linear map. Let $\mathbf{M}$ represent the adjacency matrix of the food web. The element $M_{ij}$ equals one if there is carbon transfer (predation) from species $i$ to $j$, and zero otherwise. 
        The \upd{robustness} index ($R_i$) and importance index ($I_i$) of species $i$ are computed through the following iterative map:
        \begin{equation}
            \left\{
            \begin{array}{lcl}
                R_i^{(n+1)}&=&\delta + \sum_j M_{ji}\,/\,{I_j^{(n)}}\\
                I_i^{(n+1)}&=&\delta + \sum_j M_{ij}\,/\,{R_j^{(n)}}
            \end{array}
            \right.
            R_i^{(0)} = I_i^{(0)} = 1 ~ \forall i.
            \label{eq:efc}
        \end{equation}
        Here, $R_i^{(n)}$ and $I_i^{(n)}$ denote the \upd{robustness} index and the importance index of the species $i$ at iteration $n$, respectively. The quantity $\delta$ is a regularization term that guarantees convergence. It does not affect the final ranking, as soon as it is way smaller than $\mathbf{M}$'s elements. In this paper, we set it at $10^{-3}$. The algorithm iterates until the convergence of both quantities. The first equation tells us that a species has a high \upd{robustness} index if it can absorb carbon from several sources, in particular from those with a low importance index. 
        \upd{Referring to the second equation, species with a low importance index are consumed by few species, which suggests they are harder to prey upon.}
        \upd{Conversely, the} second equation implies that the importance index of a species is high if it is consumed by several other species, particularly those with a low \upd{robustness} index. Low-\upd{robustness} index species can indeed only absorb carbon from a limited set of sources, so the extinction of any of those would put them in serious danger. For this reason, we identify vulnerability as the inverse of the \upd{robustness} index. The lower the \upd{robustness} index of a species, the more vulnerable to food web shocks it is expected to be. 

        \begin{figure}[t]
            \centering
            \includegraphics[width=1\columnwidth]{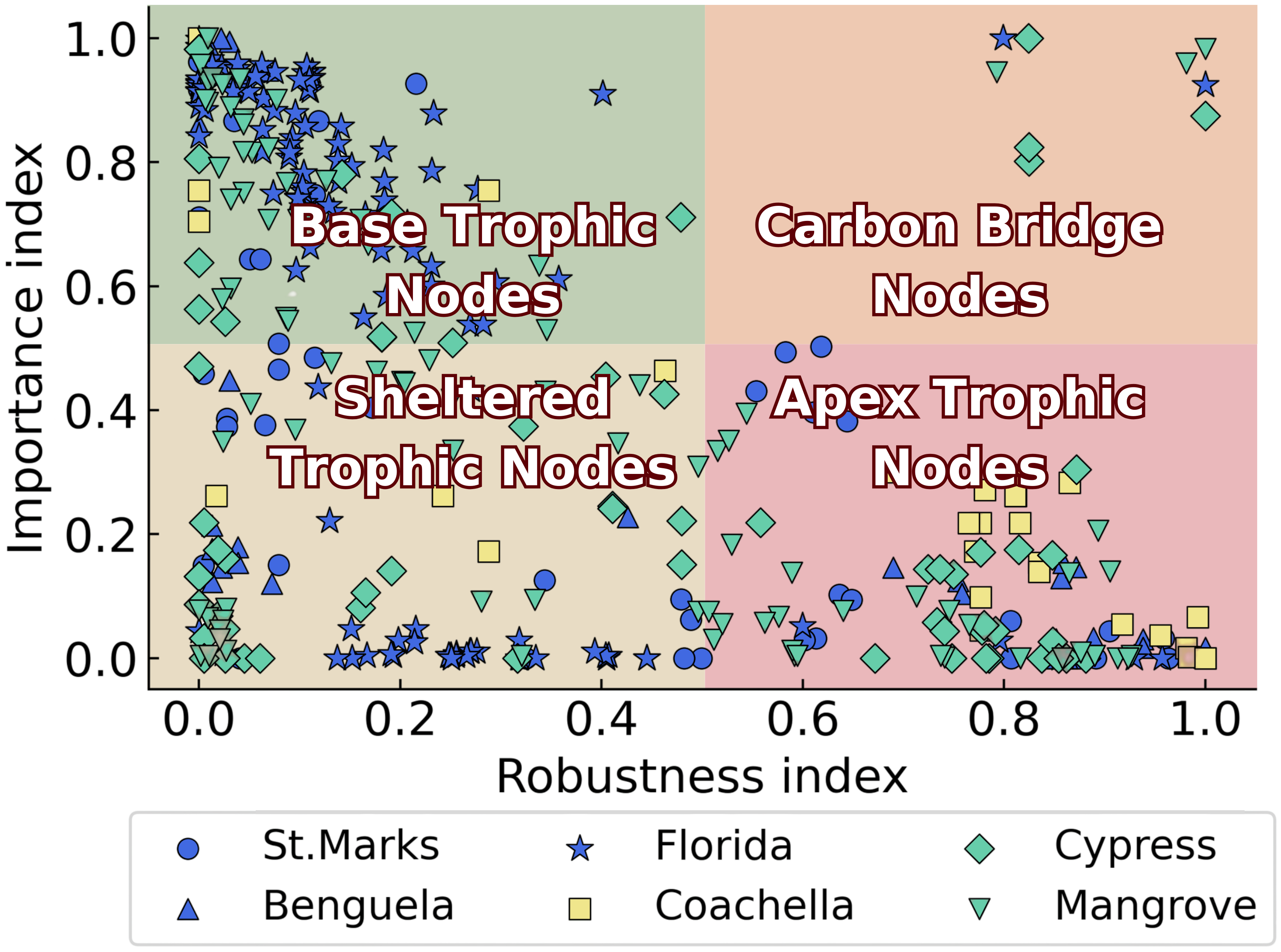}
            \caption{\textbf{\upd{Robustness} index-importance index plane.} In royal blue, the marine ecosystems; in khaki, the terrestrial one; in medium aquamarine the mixed ones. Values are log-transformed and normalized to [0,1] for cross-food web comparison. The plane is divided into four quadrants. Base Trophic Nodes (top left quadrant), Carbon Bridge Nodes (top right quadrant), Sheltered Trophic Nodes (bottom left quadrant), Apex Trophic Nodes (bottom right quadrant).}
                \label{fig:fi_plane}
        \end{figure}
        
        This approach shares notable similarities with the algorithms developed for bipartite networks in the field of Economic Complexity \cite{hidalgo2009building,tacchella2012new,servedio2018new,mazzolini2024ranking}, specifically with the algorithms of Economic Fitness and Complexity (EFC) \cite{tacchella2012new,servedio2018new}. 
        While the EFC algorithm has successfully analyzed various bipartite nested networks, including ecological systems \cite{dominguez2015ranking}, its application to mono-partite directed networks represents a novel extension \cite{servedio2025fitness}. 
        In ecological contexts, nestedness contributes significantly to ecosystem stability, biodiversity, and resilience, particularly in mutualistic networks where specialist species typically interact with generalist ones \cite{mariani2019nestedness}.
        This similarity between the two approaches is not surprising, since
        \upd{Fig.~\ref{fig:cypress}b illustrates how the directed trophic interactions of a food web can be visually represented using a bipartite layout, which can be useful for specific analytical approaches.}
        Differently from standard bipartite networks, layers contain the same set of species. The first looks at their role as predator, and the second one looks at their role as prey. Links between the two layers connect predators to their prey.

    \subsection{\upd{Robustness} index-Importance index Plane}
        \noindent
        Our algorithm allows us to place all the species within an ecosystem on the 2-dimensional space defined by the \upd{robustness} index and the importance index. Each axis captures a different aspect of species, namely their importance as prey and carbon suppliers and their prowess as predators and carbon absorbers. As an example, we show in Fig.~\ref{fig:cypress}c the \upd{robustness} index-importance index plane for the Cypress dry season food web \cite{ulanowicz1998network}. In this figure, each point corresponds to a species, different colors/shapes correspond to five different categories of species (see Methods for details), while point size is proportional to the biomass. Remarkably, species are not randomly scattered on the plane but tend to cluster following the classification. Producers and Organic Matter (dark green circles) are mostly placed on the top left corner since they are crucial carbon suppliers (high importance index) and they do not absorb carbon from other species (low \upd{robustness} index). Conversely, large predators are placed in the bottom right corner, since they are both very hard to prey on and very good at preying. Also, a large fraction of secondary consumers are located in this region. The top right corner contains species that can absorb carbon from several other species, but that, at the same time, also transfer carbon to many other species. Not surprisingly in this area we find invertebrates and detritivores or mammals like rats. Finally, the bottom left corner mostly contains herbivores/omnivores and secondary consumers. These are species such as lizards that are not central in the food chain functioning, being hard to prey on, but with little to none preying capabilities.

        In Fig.~\ref{fig:fi_plane}, the \upd{robustness} index-importance index plane for each species is shown across all food webs. The plane is then divided into four regions:
        \begin{itemize}
            \item \textbf{Base Trophic Nodes (top left quadrant)}  \\These species are characterized by poor predatory ability but are heavily preyed upon by a multitude of species. They form the foundation for energy and nutrient flow, providing a primary source of food for consumers. They are both very important and very vulnerable (low \upd{robustness} index).
            \item \textbf{Carbon Bridge Nodes (top right quadrant)}  \\These species excel at absorbing or transforming carbon (often from detrital or non-living sources) and are widely consumed by other species. They ``bridge'' carbon from lower-level or non-living pools into higher trophic levels. They are also very important, but not as vulnerable, since their high \upd{robustness} index allows them to absorb carbon from very diverse sources.
            \item \textbf{Sheltered Trophic Nodes (bottom left quadrant)}  \\These species are neither strong predators nor frequent prey; they have defenses, behaviors, or ecological niches that protect them from heavy predation. They have a low \upd{robustness} index, so high vulnerability, but their extinction would likely not cause cascades. 
            \item \textbf{Apex Trophic Nodes (bottom right quadrant)}  \\These species possess high predatory ability with low vulnerability. They regulate prey populations and structure ecosystem dynamics from the top of the food web. They have a low importance index, but also low vulnerability. 
        \end{itemize}
        Fig.~\ref{fig:fi_plane} is realized considering six different ecosystems from terrestrial, aquatic, and mixed domains. Note that we normalized the \upd{robustness} index and the importance index to compare species from different ecosystems (details in Methods). Colors denote the ecosystem type, while symbols show the different food webs. We observe a predominance of species in the Base Trophic Nodes ($42.09\%$) and Apex Trophic Nodes ($32.65\%$) quadrants, only a few Carbon Bridge Nodes ($2.30\%$) and a minority of Sheltered Tropic Nodes ($22.96\%$). We report in SI, for each ecosystem, the detailed list of all species in each quadrant. 
        \upd{It is important to stress that the robustness index and importance index only present a marginal anti-correlation (average Spearman $s = -0.60 \pm 0.18$ across the six food webs), confirming that the two quantities capture complementary species’ properties. This moderate anti-correlation is notably lower than the correlation we observe between eigenvector in-centrality and importance index (e.g., $s =0.93$ for the Cypress food web, as shown in Supplementary Information Section 3), indicating that the robustness index and importance index provide more distinct information.}

    \subsection{Cascading Extinctions and Robustness of Species}
        \noindent

        \begin{table}[t]
        \centering
        \begin{tabular}{|l|c|c|c|}
        \hline
        Food web  & In-degree (\%) & Eigenvector-in (\%) & Importance index (\%) \\ \hline
        St. Marks  & 84.8 $\pm$ 1.1 & \textbf{91.6} & 91.4 \\ \hline
        Benguela   & 77.3 $\pm$ 0.5 & \textbf{94.6} & 91.3 \\ \hline
        Coachella  & 70.9 $\pm$ 0.3 & 75.9 & \textbf{78.8} \\ \hline
        Florida    & 64.6 $\pm$ \upd{0.4} & \textbf{82.4} & \upd{82.2} \\ \hline
        Cypress    & 78.5 $\pm$ 0.4 & 81.9 & \textbf{86.9}\\ \hline
        Mangrove   & 72.5 $\pm$ 0.1 & \textbf{95.0} & 94.3 \\ \hline
        \end{tabular}
        \caption{\textbf{Comparison of mean extinction area across ecosystems using different species ranking methods.} 
        We compute the in-degree, eigenvector-in, and importance index rankings at the outset and keep them fixed throughout the simulated extinction process. Values represent means and standard deviation. For eigenvector-in and importance index, the standard deviation is smaller than the last significative digit displayed. Higher percentages indicate greater ecosystem sensitivity to species loss based on the respective ranking method. Bold values indicate the best-performing method for each ecosystem.}
        \label{table:ea_no_recomp}
        \end{table}

        \noindent
        The picture suggested by the \upd{robustness} index-importance index plane, while reasonable, needs stronger validation. In particular, we need to understand whether the two measures we introduced are correlated with the predatory capabilities of animals and their vulnerability and importance as prey. These are not easy quantities to measure, but we can indirectly assess them. 
        
        We start studying the importance index, which, as mentioned, is a measure of a species' relevance as a carbon source. High-importance index species play a major role in food webs since they represent the base of the food chain. This role can be quantified looking at the so called extinction curve of a food web. Such a curve is obtained by progressively removing species from the food web and looking at which other species get disconnected, thus remaining without any supplier. More details are reported in the Methods section. The premise is that a more accurate measure of species relevance within a food web should lead to a faster network collapse when species are removed in order of their ranking. Indeed, Eq.~(\ref{eq:efc}) implies that a species with a high importance index is preyed upon by several species of low \upd{robustness} index. Therefore, its removal would likely cause many of these species with low predatory capabilities to be unable to survive. We quantify this effect using the mean extinction area (area under the extinction curve), where higher values indicate a better ability to identify critical species within the ecosystem. Table~\ref{table:ea_no_recomp} presents the mean extinction areas following 3 different ranking strategies: in-degree, eigenvector-in centrality, and importance index. 
        We establish the initial rankings and keep them fixed throughout the simulated extinction process, whereas in the Supplementary Information, we present the scenario where rankings are recalculated after each node removal, including extinction curves plots for both cases across all food webs.
        The results reveal that both the eigenvector-in and importance index consistently outperform the degree-centrality across all ecosystems studied. Notably, eigenvector-in centrality and importance index demonstrate comparable efficacy, with each measure excelling in different ecosystems. These results validate the importance index, confirming its ability to identify the most crucial species within food webs. 

        \begin{figure*}[t]
            \centering
            \includegraphics[width=2\columnwidth]{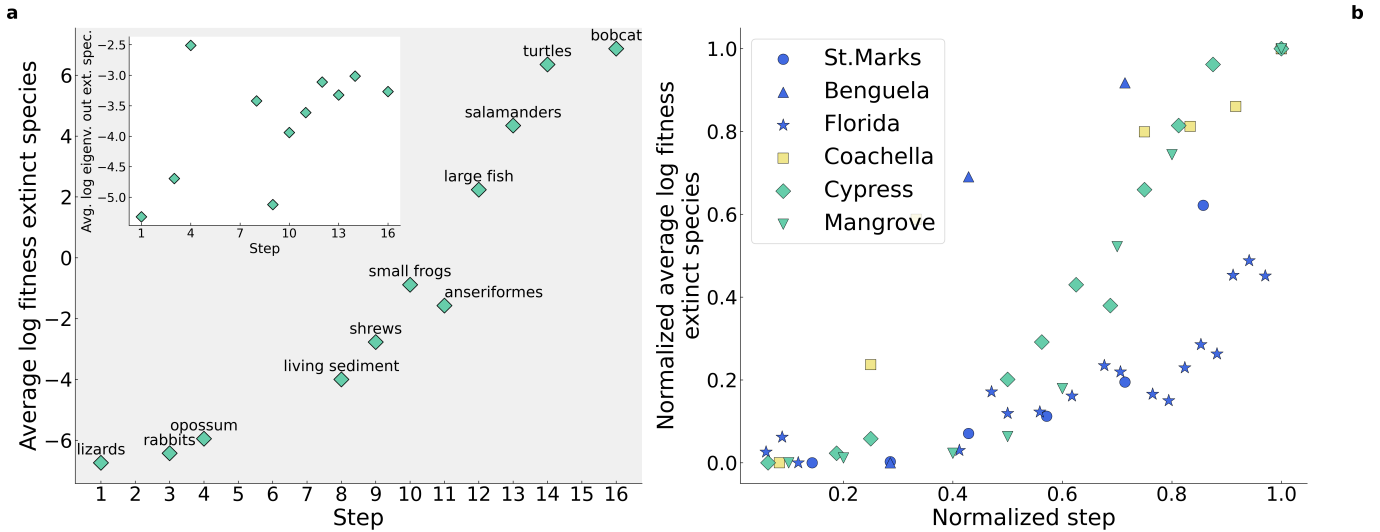}
            \caption{\textbf{Ecological robustness: Species removal patterns in diverse food webs.}
            \textbf{Left:} Cypress food web: Average log \upd{robustness} index of extinct species vs. extinction step, based on importance index ranking. Labels indicate representative species becoming extinct at each step. \textbf{Inset:} Average log eigenvector-out centrality of extinct species vs. extinction step.
            \textbf{Right:} Normalized average log \upd{robustness} index vs. normalized extinction step for multiple food webs. Values are log-transformed and scaled to [0,1] for cross-ecosystem comparison. This plot illustrates ecosystem resilience patterns and differential impacts of species loss across varied food web structures.}
            \label{fig:fit}
        \end{figure*}

        As a second step, we study the ability of the \upd{robustness} index to measure species' vulnerability. A high-\upd{robustness} index species can prey on many other organisms, including those with a low importance index, i.e., those that are very hard to prey on. As a consequence, we expect high-\upd{robustness} index species to be less affected by extinctions within the ecosystem. They are indeed very diversified and less susceptible to such events. On the other hand, low-\upd{robustness} index species are expected to be the most vulnerable and, thus, the first to get extinct during cascade events. We can measure this by looking at the average (log) \upd{robustness} index of species that get extinct at a given step of the extinction curve construction. As shown in Fig.~\ref{fig:fit}a for the cypress food web, this average increases with the step, meaning that high-\upd{robustness} index species only get extinct when most other species are already gone. We report next to each point a sample species going extinct in that step, while the inset shows the same plot, but using the average eigenvector-out centrality to measure (inverse) vulnerability. A much weaker correlation is observed. We repeat the same analysis for all the six food webs we considered, reporting the results in Fig.~\ref{fig:fit}b. Note that we normalized both the step and the \upd{robustness} index for comparison reasons. In all cases, we observe the same behavior. More quantitatively, we can compute the Spearman correlation between the removal step and the (average log) \upd{robustness} index of the removed species. The average of this quantity among all ecosystems is $0.98$, while the average of the eigenvector-out centrality is $0.72$, thus consistently lower. 
        We repeat the analysis using eigenvector-in centrality for species removal, yielding similar results that support the \upd{robustness} index as an effective measure of species' predatory skills and extinction resilience.
        In the Supplementary Information, we report details on Spearman correlation coefficients and eigenvector-in-based removals, along with additional comparisons between \upd{robustness} index, importance index, and other centrality measures.

\section{Discussion}
\noindent
    Ecosystems face intensifying threats, making it urgent to identify vulnerable species, to prevent their extinction. Current network-based approaches often reduce the complex predator-prey interactions to a single metric. While this allows us to understand which species are crucial for the functioning of a food web, it tells little to nothing about the vulnerability of species. Our method, based on a bi-dimensional non-linear map, addresses this gap, associating to each species both an \emph{importance index} and a \emph{\upd{robustness} index}. The former is linked to the susceptibility to predation and measures how crucial a species is in the food web. The latter measures predatory prowess and quantifies the robustness of species. High-importance index species, which often anchor lower trophic levels, appear critical to maintaining overall network stability, and their loss disproportionately impacts a wide range of other species. On the other hand, high-\upd{robustness} index species thrive on diverse prey pools, making them more resistant to ecosystem perturbations. Our analysis confirms this intuition, showing that the importance index effectively identifies species whose removal triggers significant co-extinctions. At the same time, the \upd{robustness} index highlights species capable of persisting until the late stages of collapse, with the most vulnerable, low-\upd{robustness} index species being the first to disappear. While eigenvector-in centrality can resemble our importance index ranking, neither eigenvector-in nor eigenvector-out centralities adequately capture the predatory strength and resilience that define a species’ \upd{robustness} index. This limitation arises from the linear nature of the eigenvector centrality, which is often incapable of capturing complex, non-linear interactions.

    By mapping species onto a \emph{\upd{robustness} index-importance index plane}, we can reveal patterns that align with known functional roles (e.g., apex predators, base trophic nodes). 
    \upd{Although expert knowledge remains crucial for fully characterizing species' functional roles, our algorithmic and data-driven approach provides a valuable, complementary layer of insight.}
    This framework holds clear promise for conservation and ecosystem management, where limited resources demand strategic prioritization. By jointly quantifying species’ importance index and \upd{robustness} index, practitioners can pinpoint ``keystone'' nodes —those whose loss risks triggering wider co-extinctions— thereby directing scarce funding and protective measures where they have the greatest impact. Particular attention can also be devoted to those species that despite a marginal role as carbon suppliers, are in severe danger due to their low \upd{robustness} index. Because our algorithmic approach draws solely on quantitative interaction data, it readily scales to large, complex food webs.
    When combined with objective risk assessment, this scalability ensures broader applicability across diverse ecosystems, ultimately supporting more targeted and cost-effective biodiversity preservation efforts.
    \upd{However, we acknowledge that the implementation of actual conservation strategies requires the consideration of numerous additional ecological, social, and economic factors beyond the scope of this network-based analysis.}
    
    While our framework provides valuable insights into species' roles in food webs, certain aspects require further exploration. 
    First, basal species are represented as individual nodes but are linked to a common root node that aggregates their shared reliance on environmental resources like light, water, and nutrients. This approach simplifies the analysis by not differentiating among their specific resource dependencies and instead focuses on their collective role as energy providers. Consequently, our analysis primarily examines the vulnerability and resilience of all other species in the food web. For basal species, however, additional factors such as Grime's classification \cite{grime2006plant}---based on competition, stress tolerance, and ruderality---may be necessary to better capture their ecological resilience and responses to cascading effects. 
    Second, species such as lizards in the Cypress food web appear peripheral in our analysis due to their limited predatory skills and resistance to predation. This could reflect biological realities, such as population constraints driven by nesting site availability rather than trophic interactions. However, it may also result from aggregation biases in network construction, where diverse prey groups are aggregated into single nodes, masking their ecological importance.
    \upd{To preliminarily assess the influence of network resolution, we conducted node aggregation experiments. Our initial findings indicate that the impact of such aggregation on our metrics of ecosystem fragility is not uniform. Aggregating multiple basal trophic nodes yielded the most pronounced effects, suggesting that the resolution at lower trophic levels could be particularly critical for accurately assessing species' roles using our framework. Further details on these aggregation experiments and their results can be found in Section 6 of the Supplementary Information. Systematic investigation into the sensitivity of our results to food web resolution, across different aggregation strategies and trophic levels, represents a valuable direction for future research.}
    \upd{Third, our analysis relies on a static and binary representation of food webs, which inherently simplifies the dynamic and often weighted nature of real ecological interactions. The species removal simulations, while useful for comparative purposes, also present a somewhat idealized view of extinction cascades, as they do not incorporate adaptive behaviors like prey switching or the continuous changes in interaction strengths that characterize natural ecosystems.}
    \upd{Finally, while our initial study focused on demonstrating the methodology across a diverse set of six existing food webs, future research could readily expand this analysis to the larger collection of currently available food web datasets to assess the generality of our findings. Subsequently, experimental validation remains a critical avenue, where collecting data on real-world ecological networks under anthropogenic disturbances could provide empirical evidence for the predictive power of the proposed framework and help quantify species' actual endangerment levels across their ecosystems.}

\section{Methods}
\subsection{Data}
\noindent
We analyzed the following six food webs:\\
\begin{description}
    \item[{Florida Bay}] 
        This network represents trophic flows through the ecosystem of Florida Bay. 
        It includes various species such as seagrass and algae producers, microfauna, macroinvertebrates, fishes, mammals, and avifauna \cite{ulanowicz1998network}.
    \item[{Cypress Dry Season}]
        This network depicts the cypress wetlands in South Florida during the dry season \cite{ulanowicz1998network}.
    \item[{Mangrove Wet Season}]
        This network illustrates the mangrove ecosystem in South Florida during the wet season \cite{ulanowicz1998network}.
    \item[St. Marks]
        This network describes a winter's \textit{Halodule wrightii} (shoal grass) community in Goose Creek Bay, St. Marks National Wildlife Refuge, Florida \cite{christian1999organizing}.
    \item[Coachella]
        This network represents the sand community in the Coachella Valley desert. The biota includes species of vascular plants, vertebrates, arachnids, microorganisms, and insects \cite{polis1991complex}.
    \item[Benguela]
        This network characterizes the Benguela Marine Ecosystem, including marine mammals and fisheries \cite{yodzis1998local}.\\[2mm]
\end{description}

\subsection{Networks Construction and Preprocessing}
\noindent
Food webs are represented as directed monopartite networks, where edges denote energy flow between species. In our binary model, an edge from node $i$ to node $j$ indicates $i$ preys upon species $j$, with no weights assigned to the links. Self-loops are included to account for cannibalism within species.
We selected six food webs for analysis and modified them to ensure irreducibility and primitivity, following the method described by \cite{allesina2009googling,allesina2004dominates,allesina2009functional}. This modification process involved:
\begin{enumerate}
    \item Adding a root node to represent the external environment.
    \item Creating links from primary producers to the root, reflecting the origin of matter in the food web.
    \item Establishing links from the root to every node, representing intrinsic matter loss and detritus accumulation, which is partially recycled back into the food web.
\end{enumerate}

We remove three nodes (Input, Output, and Respiration) from the Florida Bay, Cypress Dry Season, and Mangrove Wet Season networks to maintain consistency across datasets. 
For the St. Marks food web, only step 3 (add the links between root and all other nodes) is performed, since the first two were already present in the data. Furthermore, we manually added one missing link, comparing the data with the adjacency matrix present in the original paper  \cite{christian1999organizing}.
These preprocessing steps ensure that all networks represent comparable ecological entities.
The macroscopic properties of these modified networks, including the number of nodes, links, and network density, are detailed in the Supplementary Information. 

\subsection{\upd{Robustness} index-Importance index Plane}
\noindent
To categorize the species of the cypress food web into 5 categories, we use ChatGPT-4o (accessed on 24/10/2024). In particular, we use the following prompt, \textit{Can you classify the nodes in this food web into 5 categories?}, followed by the list of species in the Cypress food web. The SI provides complete species categorizations.

To enable cross-network comparisons, we normalize the \upd{robustness} index and importance index values for each food web using the following procedure:
\begin{enumerate}
    \item Compute the \upd{robustness} index and importance index for each species.
    \item Take the logarithm of these values.
    \item Scale the log-transformed values to the [0,1] interval using min-max normalization:
    \begin{equation*}
        x_\mathrm{normalized} = \frac{\log(x) - \min(\log(x))}{\max(\log(x)) - \min(\log(x))}.
    \end{equation*}
\end{enumerate}
This approach preserves relative differences while allowing for consistent comparison across different food webs.

\subsection{Extinction Areas}
\noindent
We implement the following species removal procedure to assess ecosystem robustness. Species are progressively removed based on rankings from specified algorithms, maintaining fixed rankings throughout the process. Cascading extinctions are monitored, with a species considered extinct when left without outgoing links or with only a self-loop. The procedure continues until all species go extinct.
An extinction curve is generated by plotting the fraction of extinct species against the number of removed species. The extinction area, calculated as the integral of this curve, measures the ecosystem vulnerability. This area equals 1 when all species go extinct after the first removal and approaches 0.5 when no secondary extinctions occur.
\upd{When ranking species for removal, multiple species can sometimes receive the same rank value (a situation known as rank degeneracy). To address this, we employ a randomized tie-breaker. When ties occur, we randomly select which of the equally ranked species to remove. To account for the variability introduced by these random choices, we average the results from 200 repetitions of the entire removal process. In each repetition, ties are broken randomly. We use this approach to compare rankings based on our importance index, eigenvector in-centrality, and in-degree centrality.}
For eigenvector-based rankings, we follow the method in \cite{allesina2009googling}: at each step, we compute the dominant eigenvector of the column-normalized adjacency matrix and remove the highest-scoring node. The dominant eigenvalue is always 1, with its associated positive eigenvector summing to 1. For eigenvector-out rankings, we use the transpose of the adjacency matrix.

\appendix
\section*{Author Contributions}
\noindent
Conceptualization: G.D.M.\ and V.D.P.S.; 
methodology: E.C., G.D.M.\ and V.D.P.S.; 
software: E.C.; 
validation: E.C., G.D.M.\ and V.D.P.S.; 
formal analysis: E.C. and G.D.M.; 
investigation: E.C. and G.D.M.; 
data curation: E.C.; 
writing---original draft preparation: E.C. and G.D.M.; 
writing---review and editing: E.C., G.D.M. and V.D.P.S.; 
visualization: E.C.\ and G.D.M.; 
supervision: G.D.M.\ and V.D.P.S.; 
project administration: V.D.P.S.
All authors have read and agreed to the published version of the manuscript.

\section*{Competing Interests}
\noindent
The authors declare no competing interests.

\section*{Data Availability}
\noindent
The edgelists and species' names for St.Marks, Coachella, and Benguela food webs were constructed by examining the adjacency matrices in their original publications \cite{christian1999organizing,polis1991complex, yodzis1998local}. 
The edgelists and species' names for the Florida Bay, Cypress Dry Season, Mangrove Wet Season food webs \cite{ulanowicz1998network} were originally available at \cite{DuBois2003}, sourced from \cite{Batagelj2006}.

All the edgelists and species' names are available in the Zenodo repository \cite{Calò_2025}.

\section*{Acknowledgments}
\noindent
This work was funded by the City of Vienna, Municipal Department 7, and the Federal Ministry of the Republic of Austria for Climate Action, Environment, Energy, Mobility, Innovation and Technology as part of the project GZ~2021-0.664.668.

We thank Fariba Karimi, Rafael Prieto-Curiel, and Andrea Perna.

\section*{Declaration of Generative AI and AI-assisted technologies in the writing process}
We acknowledge the use of LLMs for text refining. The authors reviewed and edited the content as needed and take full responsibility for the content of the publication.

\bibliography{bibliography}
\clearpage


\onecolumngrid

\appendix
\renewcommand{\thesection}{S\arabic{section}}
\section*{Supplementary Information}
\setcounter{section}{0}

\section{Network Statistics}
Table \ref{tab:network_stats} presents key statistics for each food web, including the number of nodes (\(n\)), number of links (\(m\)), and network density (\(d\)). The number of nodes includes an additional root node for each network, and the number of links considers self-loops. Network density is calculated as \(d = \frac{m}{n^2}\) for directed networks with self-loops.

\begin{table}[h!]
\centering
\begin{tabular}{|l|c|c|c|}
\hline
\textbf{Food Web}       & \textbf{\# nodes} & \textbf{\# links}  & \textbf{density (\%) }\\ \hline
St. Marks       &    49             & 275      & 11.5                \\ \hline
Benguela        &    30            & 234      & 26.0                   \\ \hline
Coachella       &    30             & 294       & 32.7                \\ \hline
Florida         & 126                & 2078 & 13.1                        \\ \hline
Cypress         & 69                 & 634 & 13.3                      \\ \hline
Mangrove        & 95                 & 1439 & 15.9                     \\ \hline
\end{tabular}
\caption{Network statistics for analyzed food webs.}
\label{tab:network_stats}
\end{table}

\section{Cypress Food Web: Chat GPT Categories}
The following are the five categories into which the species of the Cypress food web have been divided, as generated by Chat GPT.
\begin{itemize}
\item \textbf{Producers/Organic Matter}: Living POC, Living sediment, Phytoplankton, Float. vegetation, Periphyton/Macroalgae, Macrophytes, Epiphytes, Understory, Vine Leaves, Hardwoods Leaves, Cypress Leaves, Cypress Wood, HW Wood, Roots
\item \textbf{Invertebrates/Detritivores}: Crayfish, Apple Snail, Prawn, Aquatic Invertebrates, Ter. Invertebrates, Refractory Det., Liable Det., Vertebrate Det.
\item \textbf{Herbivores/Omnivores}: Small Fish, herb and omniv, Small Frogs, Tadpoles, Galliformes, White ibis, Gruiformes, Caprimulgiformes, Hummingbirds, Woodpeckers, Passeriformes onniv., Squirrels, Mice \& Rats, Rabbits, White-Tailed Deer, Armadillo
\item \textbf{Secondary Consumers}: Small Fish, prim. carniv, Large Fish, Turtles, Lizards, Snakes, Salamanders, Large Frogs, Medium Frogs, Salamander L, Pelecaniformes, Anseriformes, Vultures, Egrets, Great blue heron, Other herons, Wood stork, Passeriformes pred., Opossum, Shrews, Bats, Mink, Otter, Bobcat
\item \textbf{Large Predators}: Alligators, Kites \& Hawks, Owls, Black Bear, G. Fox, Raccoon, Florida Panther, Hogs
\end{itemize}

\section{Robustness index and Importance index}
\subsection{Quadrants Robustness index - Importance index Plane}
Species in each food web are categorized into four quadrants based on their robustness index and importance index: Base Trophic Nodes, Sheltered Trophic Nodes, Apex Trophic Nodes, and Carbon Bridge Nodes. The full species lists for each category are provided in Appendix \ref{app:species_cat}.

\subsection{Correlation with Eigenvector Centralities}
\begin{figure}[h]
            \centering
            \includegraphics[width=0.85\columnwidth]{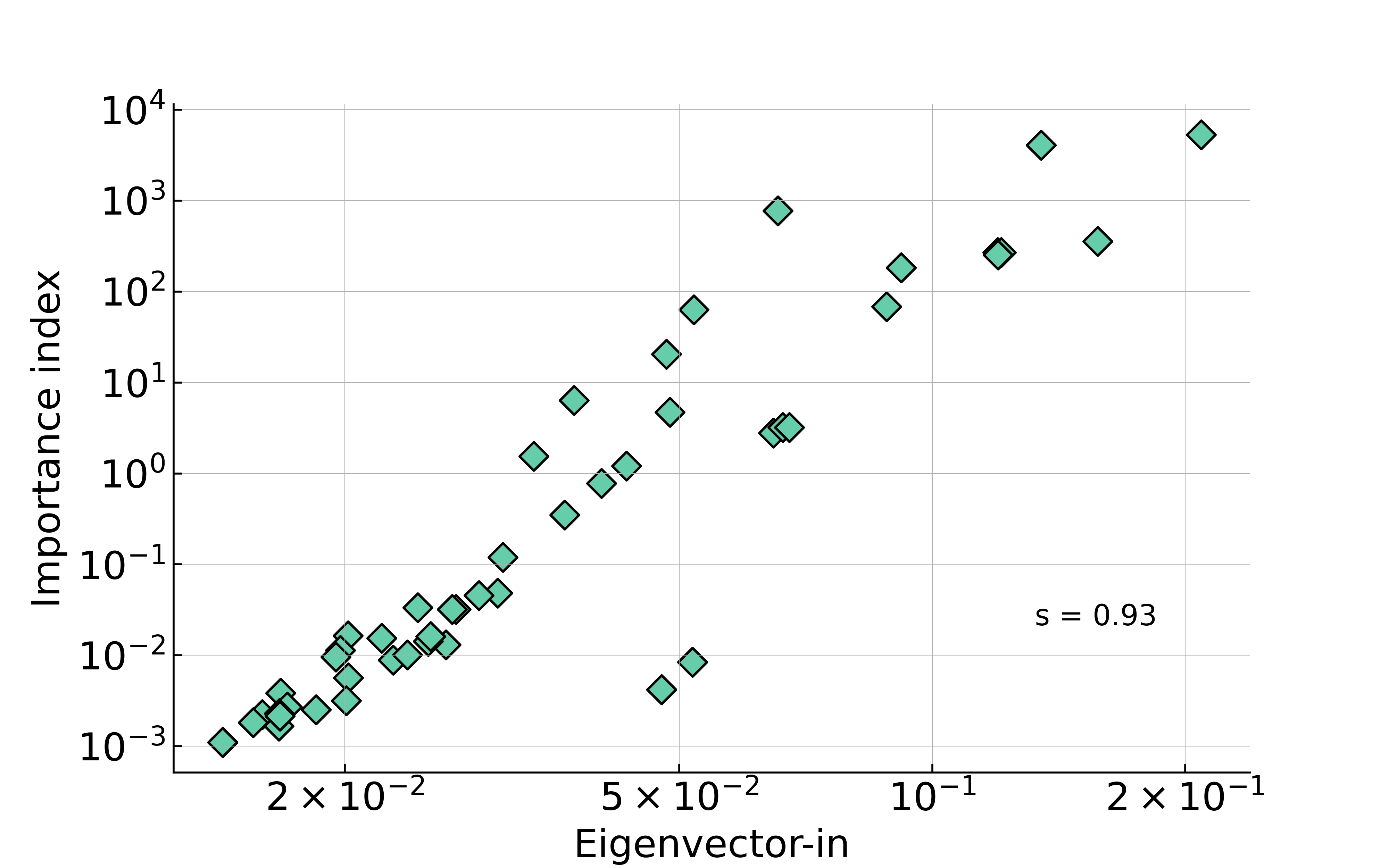}
            \caption{\textbf{Relationship between eigenvector-in centrality and importance index in the Cypress food web.}
            The Spearman correlation coefficient (s) is reported.}
            \label{fig:eig_in_importance}
        \end{figure}

\begin{figure}[h]
            \centering
            \includegraphics[width=0.85\columnwidth]{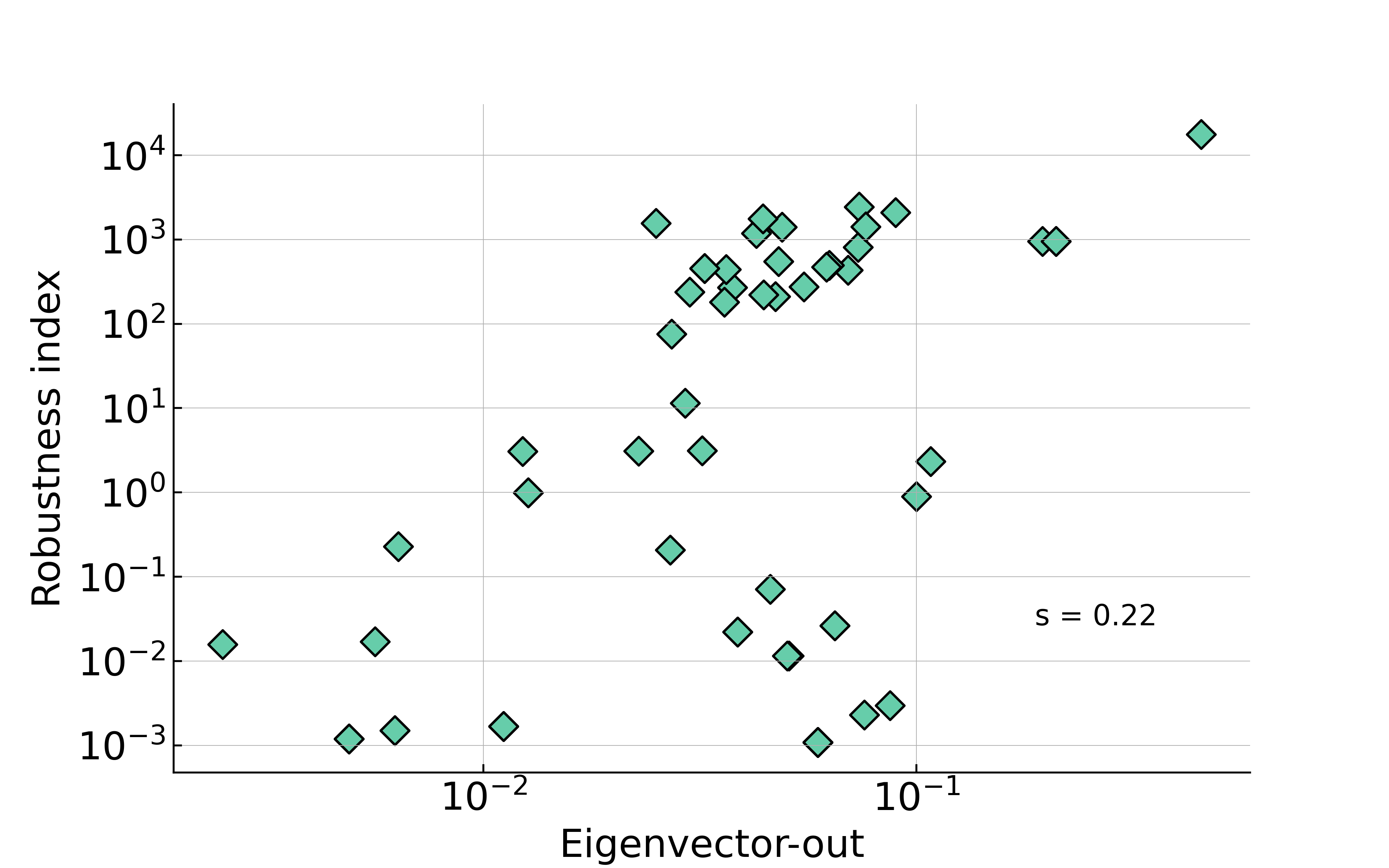}
            \caption{\textbf{Relationship between eigenvector-out centrality and robustness index in the Cypress food web.}
            The Spearman correlation coefficient (s) is reported.}
            \label{fig:fitness}
        \end{figure}

Figure \ref{fig:eig_in_importance} reveals a strong correlation between the importance index and eigenvector-in centrality (Spearman coefficient = 0.93), explaining their similar performance in extinction area analysis. Conversely, Figure \ref{fig:fitness} shows a weak correlation between the robustness index and eigenvector-out centrality (Spearman coefficient = 0.22), accounting for significant differences in identifying resilient species.

\clearpage

\section{Cascading Extinctions}
\subsection{Recomputed Rankings}
Table~\ref{table:ea_recomp} compares mean extinction areas using different ranking strategies, recomputing rankings after each species removal.
The results reveal that in this case eigenvector-in centrality generally outperforms the importance index, which in turn outperforms in-degree. Notably, for the Coachella food web, the importance index shows superior performance compared to eigenvector-in centrality.

\begin{table}[t!]
        \centering
        \begin{tabular}{|l|c|c|c|}
        \hline
        \textbf{Food Web}  & \textbf{In-degree (\%)} & \textbf{Eigenvector-in (\%)} & \textbf{Importance index (\%)} \\ \hline
        St. Marks  & 84.2 $\pm$ 1.1 & \textbf{91.5} & 87.4 \\ \hline
        Benguela   & 78.2 $\pm$ 0.3 & \textbf{94.7} & 87.9 \\ \hline
        Coachella  & 71.5 $\pm$ 0.1 & 71.6 & \textbf{79.3} \\ \hline
        Florida    & 63.5 $\pm$ 0.0 & \textbf{87.3} & 77.8 \\ \hline
        Cypress    & 78.6 $\pm$ 0.3 & \textbf{86.6} & 82.2\\ \hline
        Mangrove   & 68.1 $\pm$ 0.4 & \textbf{95.3} & 92.3 \\ \hline
        \end{tabular}
        \caption{\textbf{Comparison of mean extinction area across ecosystems using different species ranking methods.} 
        We compute the in-degree, eigenvector-in, and importance index rankings at the outset and recompute them after each step of the simulated extinction process. Values represent means and standard deviation. The standard deviation for the eigenvector-in and importance index is smaller than the last significative digit displayed (except for the eigenvector-in in the Coachella food web, for which it is 1.0). Higher percentages indicate greater ecosystem sensitivity to species loss based on the respective ranking method. Bold values indicate the best-performing method for each ecosystem.}
        \label{table:ea_recomp}
\end{table}

\begin{table}[t]
        \centering
        \begin{tabular}{|l|c|c|c|}
        \hline
        \textbf{Food Web}  & \textbf{In-degree (\%)} & \textbf{Eigenvector-in (\%)} & \textbf{Importance index (\%)} \\ \hline
        St. Marks  & 84.8 $\pm$ 1.1 & \textbf{91.6} & 91.4 \\ \hline
        Benguela   & 78.2 $\pm$ 0.3 & \textbf{94.7} & 91.3 \\ \hline
        Coachella  & 71.5 $\pm$ 0.1 & 75.9 & \textbf{79.3} \\ \hline
        Florida    & 64.6 $\pm$ 0.3 & \textbf{87.3} & 82.6 \\ \hline
        Cypress    & 78.6 $\pm$ 0.3 & 86.6 & \textbf{86.9}\\ \hline
        Mangrove   & 72.5 $\pm$ 0.1 & \textbf{95.3} & 94.3 \\ \hline
        \end{tabular}
        \caption{\textbf{Comparison of maximum mean extinction areas across ecosystems.} For each food web and ranking method (in-degree, eigenvector-in, and importance index), the table presents the higher mean extinction area obtained from either the fixed or recomputed ranking strategy.
        }
        \label{table:best}
        \end{table}

\subsection{Recomputed vs. Fixed Rankings}
Table \ref{table:best} shows the maximum mean extinction areas for each food web and ranking method (in-degree, eigenvector-in, importance index). The reported value represents the better result between the fixed and recomputed ranking strategies.

\begin{figure}[t]
            \centering
            \includegraphics[width=0.8\columnwidth]{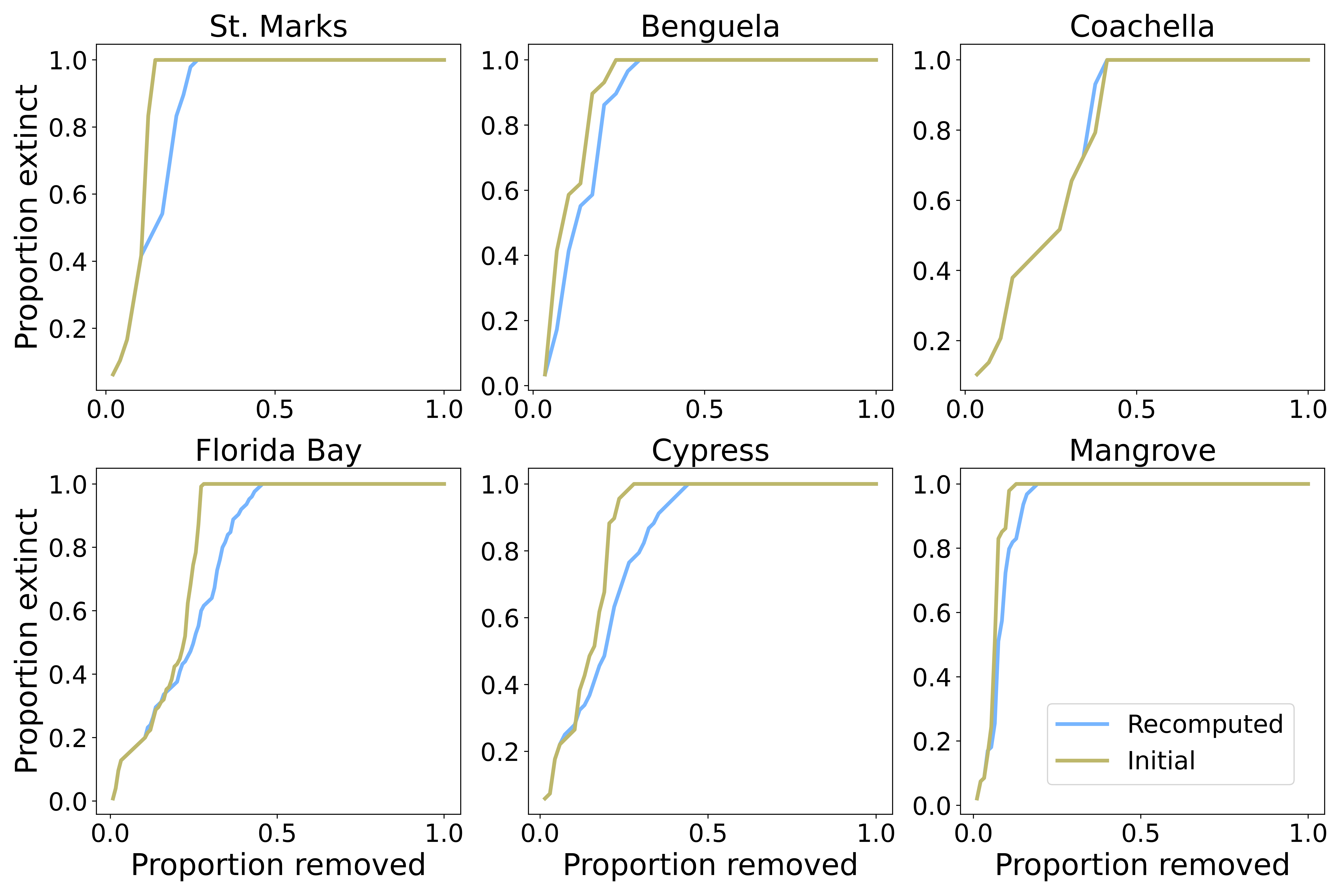}
            \caption{\textbf{Comparative extinction curves across food webs.} Extinction trajectories for all six food webs, illustrating species loss under two removal strategies: dynamically recomputed importance index ranking and fixed initial importance index ranking. Each subplot depicts the proportion of extinct species versus the proportion of species removed, highlighting differences between adaptive and static removal processes.}
            \label{fig:ext_cur}
        \end{figure}

In Figure \ref{fig:ext_cur}, we compare the extinction curves for all food webs under two removal strategies (recomputed and fixed initial ranking) for importance index ranking. Coachella is the only food web where recomputing outperforms the fixed ranking.

        \begin{table}[h!]
        \centering
        \begin{tabular}{|l|c|c|}
        \hline
        \textbf{Food Web}   &  \textbf{Robustness index (\%)} & \textbf{Eigenvector-out (\%)}\\ \hline
        St. Marks   & 100.00 & 92.86 \\ \hline
        Benguela    & 100.00 & 80.00\\ \hline
        Coachella   & 100.00 & 60.71  \\ \hline
        Florida    & 91.93  & 55.09  \\ \hline
        Cypress   & 99.09 & 54.55 \\ \hline
        Mangrove   & 100.00 & 90.48 \\ \hline
        \end{tabular}
        \caption{\textbf{Spearman rank correlation coefficients (\%) between removal steps and ecosystem metrics:} average log robustness index and average log eigenvector-out of extinct species. Removals were performed following the importance index ranking.}
        \label{table:spearmann}
\end{table}

\newpage
\section{Robustness of Species}

Table \ref{table:spearmann} presents the Spearman rank correlation coefficients between removal steps and two ecosystem metrics: average log robustness index and average log eigenvector-out centrality of extinct species. These coefficients were calculated based on species removals following the importance index ranking. The results demonstrate that the robustness index consistently outperforms eigenvector-out centrality in identifying resilient species across all food webs. 

\begin{figure}[t]
            \centering
            \includegraphics[width=1\columnwidth]{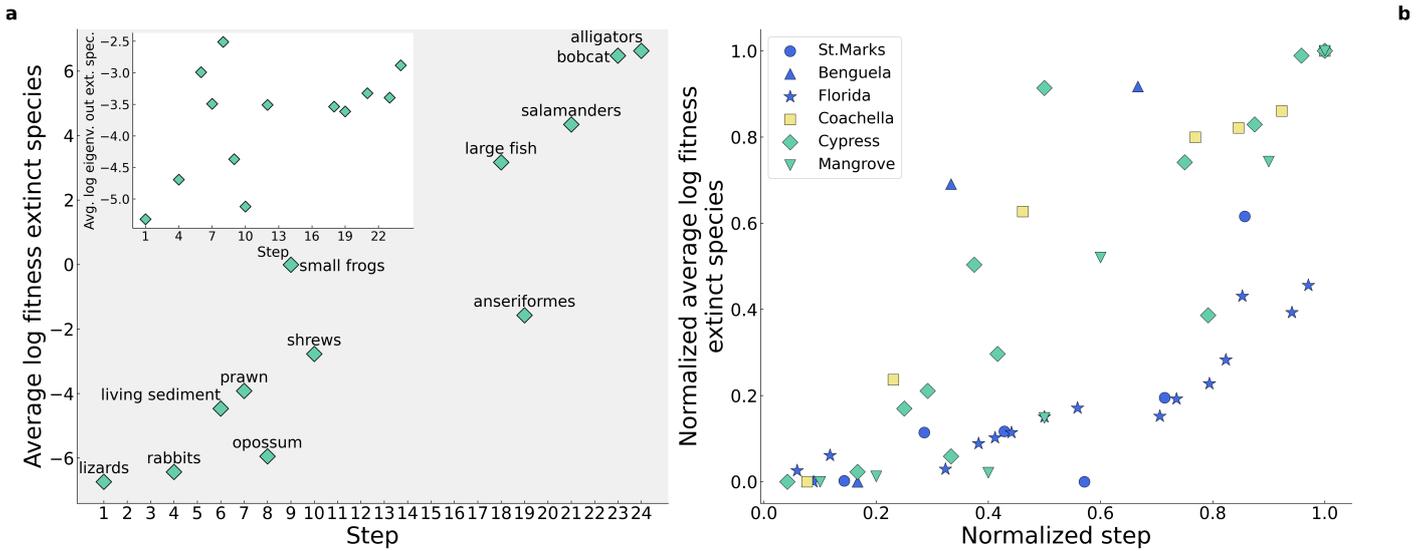}
            \caption{\textbf{Ecological robustness: Species removal patterns in diverse food webs.}
            \textbf{Left:} Cypress food web: Average log robustness index of extinct species vs. extinction step, based on eigenvector-in ranking. Labels indicate representative species becoming extinct at each step. \textbf{Inset:} Average log eigenvector-out centrality of extinct species vs. extinction step.
            \textbf{Right:} Normalized average log robustness index vs. normalized extinction step for multiple food webs. Values are log-transformed and scaled to [0,1] for cross-ecosystem comparison. This plot illustrates ecosystem resilience patterns and differential impacts of species loss across varied food web structures.}
            \label{fig:resilience}
        \end{figure}

This superior performance of the robustness index holds true even when species removals are based on eigenvector-in centrality, as illustrated in Figure \ref{fig:resilience}a for the Cypress food web.
Figure \ref{fig:resilience}b extends this analysis to all six food webs, showcasing normalized average log robustness index versus normalized extinction steps. This comparison reveals consistent patterns of ecosystem resilience across diverse food web structures.

        \begin{figure}[h!]
            \centering
            \includegraphics[width=0.8\columnwidth]{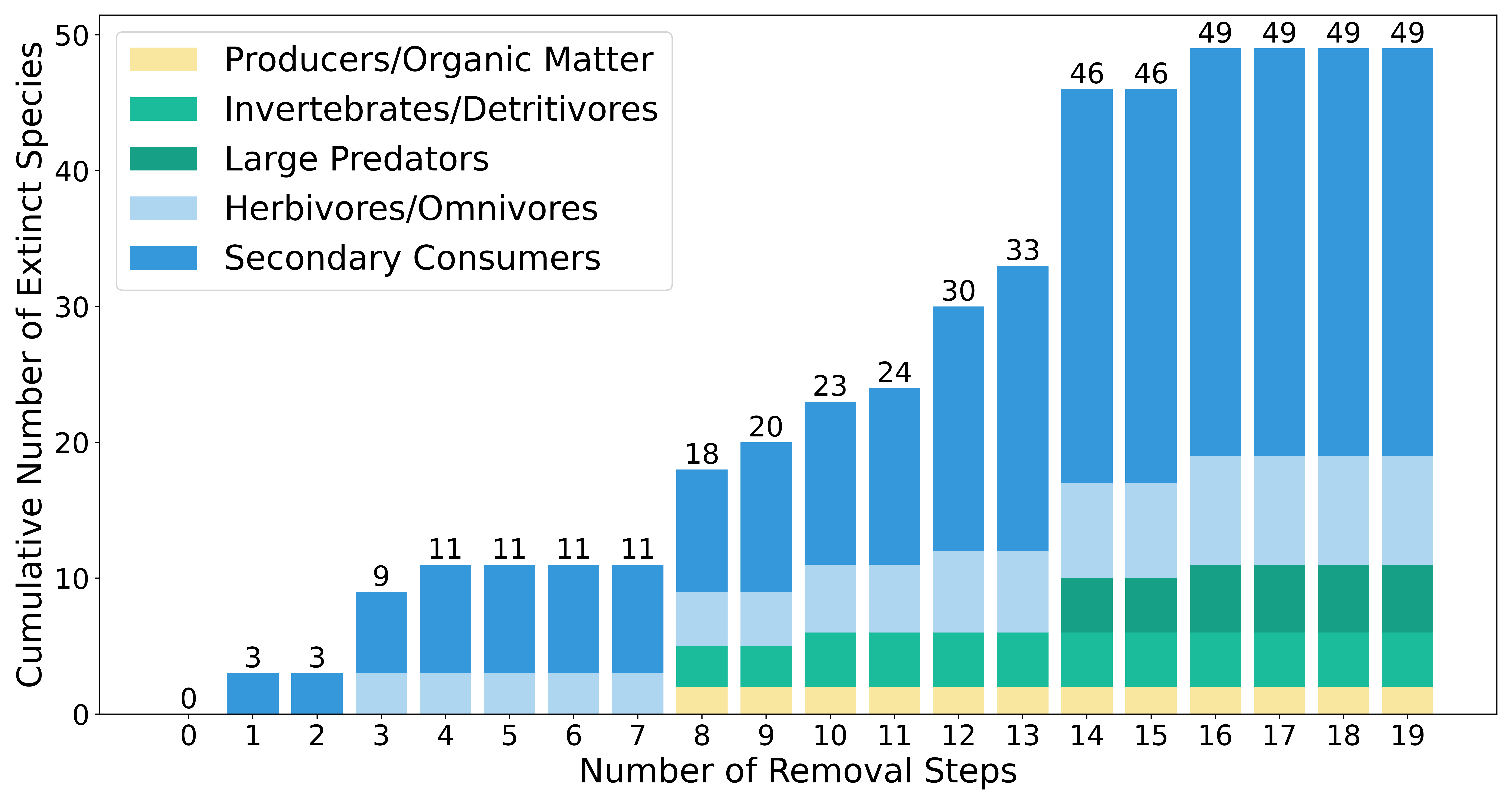}
            \caption{\textbf{Cumulative species extinctions by functional category in the Cypress food web, based on the importance index ranking.}
            Stacked bar chart illustrating the progressive accumulation of extinct species at each removal step. Colors represent different functional categories. Bar heights remain constant when no new extinctions occur, reflecting the cumulative nature of the data. This visualization demonstrates the differential impact of species removals on various functional groups within the ecosystem.}
            \label{fig:bars}
        \end{figure}


In the specific case of the Cypress food web, Figure \ref{fig:bars} provides additional insights into the extinction dynamics. 
The visualization reveals that large predators, primarily Apex Trophic Nodes, persist until later removal stages, consistent with their high robustness index scores and broad prey bases.

\section{Aggregation Analysis}
In the Florida Bay food web, we aggregated eight phytoplankton species ($2\mu$m Spherical Phytoplankt, Synedococcus, Oscillatoria, Small Diatoms ($<20\mu$m), Big Diatoms ($>20\mu$m), Dinoflagellates, Other Phytoplankton, Benthic Phytoplankton) into a single ``Phytoplankton" node. We also aggregated two heron categories (Big Herons \& Egrets, Small Herons \& Egrets) into a single ``Herons \& Egrets" node, three turtle species (Loggerhead Turtle, Green Turtle, Hawksbill Turtle) into ``Turtles", and three killifish species (Other Killifish, Goldspotted killifish, Rainwater killifish) into ``Killifish". 
We then examined the effect of these individual aggregations on the total extinction area resulting from simulated species removals and cascading effects. Aggregating the eight phytoplankton species led to the most substantial increase in the extinction area (eigenvector centrality-based area to 91.0, importance-based area to 90.1). The aggregation of Herons \& Egrets resulted in a minor change. Aggregating turtles led to a slight decrease in the eigenvector centrality-based extinction area (to 82.1) and a minor increase in the importance-based area (to 82.4). Aggregating killifish resulted in a slight decrease in both the eigenvector centrality-based extinction area (to 82.3) and the importance-based area (to 82.1).

\clearpage
\appendix
\section{Species Categorization}
\label{app:species_cat}

Tables \ref{tab:food_web_species_1} and \ref{tab:food_web_species_2} present the species categorization for each food web based on their position in the robustness index-importance index plane. 
The Carbon Bridge Nodes for each relevant food web are as follows:

\begin{itemize}
    \item Florida Bay: Benthic POC and Water POC.
    \item Cypress: Terrestrial Invertebrates, Liable Detritus, Refractory Detritus, and Vertebrate Detritus.
    \item Mangrove: POC, INSCT, and C in SED.
\end{itemize}
        
\begin{table}[h]
\begin{adjustbox}{center}
\footnotesize
\begin{tabular}{|p{1.5cm}|p{4cm}|p{4cm}|p{4.5cm}|}
\hline
\textbf{Food Web} & \textbf{Base Trophic Nodes} & \textbf{Sheltered Trophic Nodes} & \textbf{Apex Trophic Nodes} \\
\hline
St. Marks & Benthic bacteria, Microfauna, Meiofauna, Bacterioplankton, Microprotozoa, Epiphyte-grazing amphipods, Suspension-feeding molluscs, Deposit-feeding peracaridan crustacean, Deposit-feeding gastropods, Other gastropods, Deposit-feeding polychaetes, Zooplankton, Halodule, Micro-epiphytes, Macro-epiphytes, Benthic algae, Phytoplankton, Detritus & Hermit crabs, Spider crabs (herbivores), Isopods, Brittle stars, Herbivorous shrimps, Tonguefish, Sheepshead minnow, Epiphyte-grazing gastropods, Suspension-feeding polychaetes & Omnivorous crabs, Blue crabs, Predatory shrimps, Catfish and stingrays, Gulf flounder and needlefish, Southern hake and sea robins, Atlantic silverside and bay anchovies, Killifishes, Gobies and blennies, Pipefish and seahorses, Red drum, Predatory gastropods, Predatory polychaetes, Benthos-eating birds, Fish-eating birds, Fish and crustacean-eating birds, Gulls, Raptors, Herbivorous ducks \\
\hline
Benguela & Phytoplankton, Benthic filter-feeders, Bacteria, Microzooplankton, Mesozooplankton, Macrozooplankton & Benthic carnivores, Gelatinous zooplankton, Anchovy, Pilchard, Round herring, Lightfish, Lanternfish, Goby & Other pelagics, Horse mackerel, Chub mackerel, Other groundfish, Hakes, Squid, Tunas, Snoek, Kob, Yellowtail, Geelbek, Whales \& dolphins, Birds, Seals, Sharks \\
\hline
Coachella & plants/plant products, detritus, carrion, soil microbes & soil microarthropods and nematodes, surface arthropod detritivores, surface arthropod herbivores, herbivorous mammals and reptiles & soil micropredators, soil macroarthropods, soil macroarthropod predators, small arthropod predators, medium arthropod predators, facultative arthropod predators, life-history arthropod omnivore, spider parasitoids, primary parasitoids, hyperparasitoids, facultative hyperparasitoids, primarily herbivorous mammals and birds, small omnivorous mammals and birds, predaceous mammals and birds, arthropodivorous snakes, primarily arthropodivorous lizards, primarily carnivorous lizards, primarily carnivorous snakes, large primarily predaceous birds, large primarily predaceous mammals, golden eagle \\
\hline
\end{tabular}
\end{adjustbox}
\caption{Species categorization for St. Marks, Benguela, and Coachella food webs based on their trophic roles.}
\label{tab:food_web_species_1}
\end{table}

\begin{table}[h]
\begin{adjustbox}{center}
\footnotesize
\begin{tabular}{|p{1.5cm}|p{6.5cm}|p{3.7cm}|p{3.5cm}|}
\hline
\textbf{Food Web} & \textbf{Base Trophic Nodes} & \textbf{Sheltered Trophic Nodes} & \textbf{Apex Trophic Nodes} \\
\hline
Florida Bay & $2\mu$m Spherical Phytoplankt, Synedococcus, Oscillatoria, Small Diatoms ($<20\mu$m), Big Diatoms ($>20\mu$m), Dinoflagellates, Other Phytoplankton, Benthic Phytoplankton, Thalassia, Halodule, Syringodium, Drift Algae, Epiphytes, Water Flagellates, Water Cilitaes, Acartia Tonsa, Oithona nana, Paracalanus, Other Copepoda, Meroplankton, Other Zooplankton, Benthic Flagellates, Benthic Ciliates, Meiofauna, Sponges, Coral, Other Cnidaridae, Echinoderma, Bivalves, Detritivorous Gastropods, Epiphytic Gastropods, Predatory Gastropods, Detritivorous Polychaetes, Predatory Polychaetes, Suspension Feeding Polych, Macrobenthos, Benthic Crustaceans, Detritivorous Amphipods, Herbivorous Amphipods, Isopods, Herbivorous Shrimp, Predatory Shrimp, Pink Shrimp, Thor Floridanus, Lobster, Detritivorous Crabs, Omnivorous Crabs, Predatory Crabs, Callinectus sapidus, Stone Crab, Sardines, Anchovy, Bay Anchovy, Lizardfish, Catfish, Eels, Toadfish, Brotalus, Halfbeaks, Needlefish, Other Killifish, Goldspotted killifish, Rainwater killifish, Sailfin Molly, Silverside, Other Horsefish, Gulf Pipefish, Dwarf Seahorse, Other Snapper, Mojarra, Grunt, Pinfish, Scianids, Parrotfish, Mullet, Blennies, Code Goby, Clown Goby, Flatfish, Filefishes, Puffer, Other Pelagic Fishes, Other Demersal Fishes & Roots, Sharks, Rays, Bonefish, Snook, Grouper, Jacks, Pompano, Gray Snapper, Porgy, Spotted Seatrout, Red Drum, Spadefish, Mackerel, Big Herons \& Egrets, Small Herons \& Egrets, Ibis, Roseate Spoonbill, Herbivorous Ducks, Omnivorous Ducks, Gruiformes, Small Shorebirds, Gulls \& Terns, Kingfisher, Loggerhead Turtle, Green Turtle, Hawksbill Turtle, Manatee, DOC & Tarpon, Barracuda, Loon, Greeb, Pelican, Comorant, Predatory Ducks, Raptors, Crocodiles, Dolphin \\
\hline
Cypress & Living POC, Living sediment, Phytoplankton, Float. vegetation, Periphyton/Macroalgae, Macrophytes, Epiphytes, Understory, Hardwoods Leaves, Crayfish, Apple Snail, Prawn, Aquatic Invertebrates, Small Fish herb and omniv, Passeriformes onniv. & Vine Leaves, Cypress Leaves, Cypress Wood, HW Wood, Roots, Lizards, Medium Frogs, Small Frogs, Salamander L, Tadpoles, Anseriformes, Vultures, Galliformes, Caprimulgiformes, Hummingbirds, Woodpeckers, Passeriformes pred., Opossum, Shrews, Bats, Rabbits, White-Tailed Deer & Small Fish prim. carniv, Large Fish, Alligators, Turtles, Snakes, Salamanders, Large Frogs, Pelecaniformes, Kites \& Hawks, Egrets, Great blue heron, Other herons, Wood stork, White ibis, Gruiformes, Owls, Black Bear, G. Fox, Raccoon, Mink, Otter, Florida Panther, Bobcat, Squirrels, Hogs, Armadillo \\
\hline
Mangrove & PHY, OTH. PP, LEAF, MICR. HO, ZOOPL., BACT.SED., FLA. SED., CIL. SED., MEIOF., MERO, EPIFN, POLY, AGAST, BVLVS, MBENTH, SCRUST, AMPHI, PENAID, CARID, OSHMP, JLOBST, OCRAB, TCRAB, DCRAB, PCRAB, FWINV, LARV, SPIDR, HERR, ANCH, KILLI, POEC, HRSE, SLVR, FWFSH, MOJA, GOBY, DOC  & WOOD, ROOT, TGAST, RAYS, EFISH, PIN, MULL, LZRD, VULT, C \& C, WOODP, PASSOMN, PASSPERD, RABT, SQIUR, M \& R, DERS, MANA & SHRK, TARP, NEED, SNOOK, CUDA, SNAP, SCIAE, OFISH, TURT, SNKS, COCO, AMPH, L \& G, PELC, CORM, BH \& E, SE \& E, IBIS, DUCK, DUCK, DUCK, K \& H, MRAPT, GUIF, SSBIRDS, G \& T, OWLS, OPSU, FOX, BEAR, RACO, M \& O, CATS, DOLP \\
\hline
\end{tabular}
\end{adjustbox}
\caption{Species categorization for Florida Bay, Cypress, and Mangrove food webs based on their trophic roles.}
\label{tab:food_web_species_2}
\end{table}

\end{document}